\documentclass[aps,pra,showpacs,groupedaddress,twocolumn,preprintnumbers,amsmath,amssymb]{revtex4}

\usepackage{graphicx}
\usepackage{dcolumn}
\usepackage{bm}
\usepackage{braket}

\begin{document}

\title{Bouncing motion and penetration dynamics in multicomponent Bose-Einstein condensates}

\author{Yujiro Eto$^{1}$}
\author{Masahiro Takahashi$^{1}$}
\author{Keita Nabeta$^{1}$}
\author{Ryotaro Okada$^{1}$}
\author{Masaya Kunimi$^{2}$}
\author{Hiroki Saito$^{2}$}
\author{Takuya Hirano$^{1}$}
\affiliation{%
$^{1}$Department of Physics, Gakushuin University, Toshima, Tokyo 171-8588, Japan\\
$^{2}$Department of Engineering Science, University of Electro-Communications, Chofu, Tokyo 182-8585, Japan}

\date{\today}
             
\begin{abstract}
We investigate dynamic properties of bouncing and penetration in colliding binary and ternary Bose-Einstein condensates comprised of different Zeeman or hyperfine states of $^{87}$Rb.
Through the application of magnetic field gradient pulses, two- or three-component condensates in an optical trap are spatially separated and then made to collide.
The subsequent evolutions are classified into two categories: repeated bouncing motion and mutual penetration after damped bounces.
We experimentally observed mutual penetration for immiscible condensates, bouncing between miscible condensates, and domain formation for miscible condensates.  
From numerical simulations of the Gross-Pitaevskii equation,
we find that the penetration time can be tuned by slightly changing the atomic interaction strengths.
\end{abstract}

\pacs{05.30.Jp, 03.75.Kk, 03.75.Mn, 67.85.Hj}
\maketitle

Multicomponent Bose-Einstein condensates (BECs) in dilute atomic gases are an attractive system for studying hydrodynamics of multicomponent quantum fluids owing to their unprecedented controllability.
One of the significant properties characterizing multicomponent fluid systems is their miscibility; different fluids are either mutually miscible or phase separation occurs.
In multicomponent BECs, the miscibility is determined by inter- and intra-species atomic interaction strengths \cite{Timmermans98}
and, importantly, they can be experimentally controlled using Feshbach resonances \cite{Inouye98, Thalhammer08} and Rabi coupling \cite{Nicklas11,Nicklas14}. 
Multicomponent BECs with various degrees of miscibility are also available by choosing the internal states \cite{Kurn98}  or atomic species \cite{Modugno02}. 
Employing such adjustability and selectability, intriguing phenomena that depend on the degree of miscibility have been experimentally observed, e.g., soliton generation in a counterflow of miscible fluids \cite{Hamner11, Hoefer11}, quantum tunneling across spin domains \cite{Kurn99} and suppression of relative flow by multiple domains \cite{Eto15} in immiscible systems.
Theoretically, quantum turbulence in a counterflow of miscible BECs \cite{Takeuchi10PRL,Fujimoto12} and pattern formation by instabilities at interfaces in immiscible BECs \cite{Sasaki09,Takeuchi10PRB,Bezett10,Suzuki10,Kadokura12} have been predicted.

The condition for miscibility in multicomponent BECs is determined by linear stability analysis of a static or steady state, and is not naively applicable to dynamical situations.
Let us consider a situation in which two wave packets of different BECs collide with each other.
One may think that the two wave packets pass through each other without much reflection for miscible BECs, while for immiscible BECs they do not.
However, we will show that these simple predictions from the miscibility do not apply to a highly dynamic situation.

In this Rapid Communication, we generate multicomponent BECs with various degrees of miscibility by utilizing the rich spin degrees of freedom of the $^{87}$Rb atom,
and investigate the dynamical properties of bouncing and penetration in colliding binary and ternary BECs in an optical trap.
We observed various dynamics, including bouncing between miscible BECs and mutual penetration of immiscible BECs, which seems counterintuitive at first glance.
In miscible and weakly immiscible binary and ternary systems, after a few bounces, BECs mutually penetrate and create the domain structure.
In contrast, in the case of a relatively strongly immiscible system, binary BECs continue to bounce.
Such repetitive bouncing motion between atoms has been observed only in a Tonks-Girardeau gas \cite{Kinoshita06}, a Fermi gas \cite{Sommer11}, and matter-wave solitons \cite{Nguyen14}.
In addition, numerical simulations of the Gross-Pitaevskii (GP) equation suggest that the penetrability and penetration time between BECs can be tuned by slightly changing the atomic interaction strength.


\begin{figure}[t]
\includegraphics[width=8.3cm]{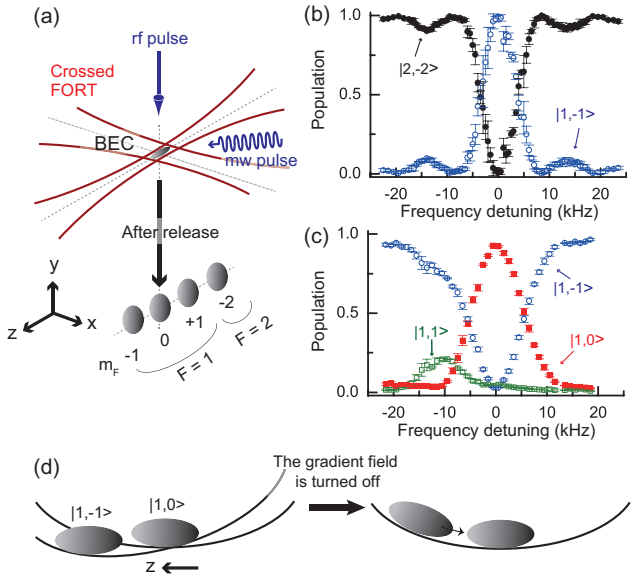}
\caption{
(color online) (a) Schematic illustration of the experimental setup.
The BEC is confined in the crossed FORT.
The multicomponent BECs are prepared using a mw pulse or an rf pulse.
After the application of a magnetic field gradient along the $z$-direction during $T_{\mathrm{sep}} =$ 60 ms and free evolution of $T_{\mathrm{evo}}$, 
the atoms are released from the FORT and then the $m_{F}$ components are separated by the SG method.
(b) Hyperfine spectroscopy.
The mw pulse with a 100 $\mu$s duration is irradiated to the BECs in the $\ket{2,-2}$ state.
Populations after the application of the mw pulse are plotted as a function of the frequency of the mw pulse.
(c) Zeeman spectroscopy.
The BECs are initially populated in $\ket{1,-1}$ state.
The shape of the applied rf pulse is a Gaussian function with a standard deviation of $23$ $\mu$s.
The small error bars in (b) and (c) enable accurate preparation of multicomponent BECs.
(d) An example of the procedure used to collide two-component BECs in an optical trap.
Two-component BECs comprised of $\ket{1,0}$ and $\ket{1,-1}$ are separated by a magnetic field gradient (left).
They collide after the magnetic field gradient is turned off (right). 
}
\label{f:schematic}
\end{figure}

\begin{table}[b]
\caption{Degrees of miscibility in binary BECs, $\Delta_{\rm a} = a_{ij} - \sqrt{a_{ii}a_{jj}}$.
Here $a_{1,2}$ represents the interspecies scattering length in units of the Bohr radius, $a_{\rm B}$.
$a_{1,1}$ and $a_{2,2}$ are intraspecies scattering lengths.
$\Delta_{\rm a} <0$ ($\Delta_{\rm a} > 0$) indicates that the ground state is miscible (immiscible).
The values of scattering lengths are obtained from Refs. \cite{Hoefer11, Widera06}.
}
\begin{ruledtabular}
\begin{tabular}{lllll}
\textbf{Binary BECs} & \textbf{$\Delta_{\rm a}$}  & \textbf{$a_{1,2}$} & \textbf{$a_{1,1}$} & \textbf{$a_{2,2}$}\\
$\ket{2, -2}$, $\ket{1,-1}$ & -0.70$a_{\rm B}$ & 98.98$a_{\rm B}$ & 98.98$a_{\rm B}$ & 100.40$a_{\rm B}$ \\
$\ket{1, 0}$, $\ket{1,-1}$ & 0.46$a_{\rm B}$ &101.09$a_{\rm B}$ & 100.86$a_{\rm B}$ & 100.40$a_{\rm B}$ \\
$\ket{1, 1}$, $\ket{1,-1}$ & 0.92$a_{\rm B}$ & 101.32$a_{\rm B}$ & 100.40$a_{\rm B}$ & 100.40$a_{\rm B}$ \\
\end{tabular}
\end{ruledtabular}
\end{table}

The outline of our experimental setup is shown in Fig. 1(a). 
We produce an $^{87}$Rb BEC containing $3 \times10^5$ atoms in the $\ket{ F = 2, m_{F} = -2 } = \ket{2,-2}$ state in a crossed far-off-resonant optical dipole trap (FORT) with axial and radial frequencies of $\omega_z / (2\pi) = 25$ Hz and $\omega_r / (2\pi) =135$ Hz (see Refs. \cite{Eto15,Eto13APEX} for a more detailed description).
In this experiment, we use three pairs of binary BECs with different degrees of miscibility, $\Delta_{\rm a}$, which are listed in Table I.
Three $m_{F}$ states in $F = 1$ are used as a ternary system.
These binary and ternary BECs, equally populated in each level, are prepared by applying resonant microwave (mw) and radio frequency (rf) pulses to the BEC in the $\ket{2,-2}$ state.
The resonant frequencies between $\ket{2, -2}$ and $\ket{1,-1}$, between $\ket{1, -1}$ and $\ket{1,0}$, and between $\ket{1, -1}$ and $\ket{1,1}$ are found by performing hyperfine spectroscopy [Fig. 1 (b)] and Zeeman spectroscopy [Fig. 1(c)].
The bias magnetic field along the $z$-direction is estimated to be $11.599$ G from these spectroscopic measurements.

In order to spatially separate the BECs, we apply a $m_{F}$-dependent potential gradient along the $z$-direction using a magnetic field gradient pulse with a duration of $T_{\mathrm{sep}} =$ $60$ ms [left side of Fig. 1(d)], whose strength is $900$ mG/cm except for the pair of $\ket{2, -2}$ and $\ket{1,-1}$.
The magnetic field gradient is reduced to 600 mG/cm for the $\ket{2, -2}$ component in order to avoid the escape of atoms from a shallow optical trap.
The magnetic field gradient is then turned off and the system evolves in the FORT for an amount of time $T_{\mathrm{evo}}$  [right side of Fig. 1(d)].
After  $T_{\mathrm{evo}}$, the BECs are released from the FORT, and each $m_{F}$ component is imaged by the Stern-Gerlach (SG) method with a time-of-flight (TOF) of 15 ms.
In order to observe the time evolution, we repeat the above procedure with different values of $T_{\mathrm{evo}}$.

\begin{figure}[t]
\includegraphics[width=8cm]{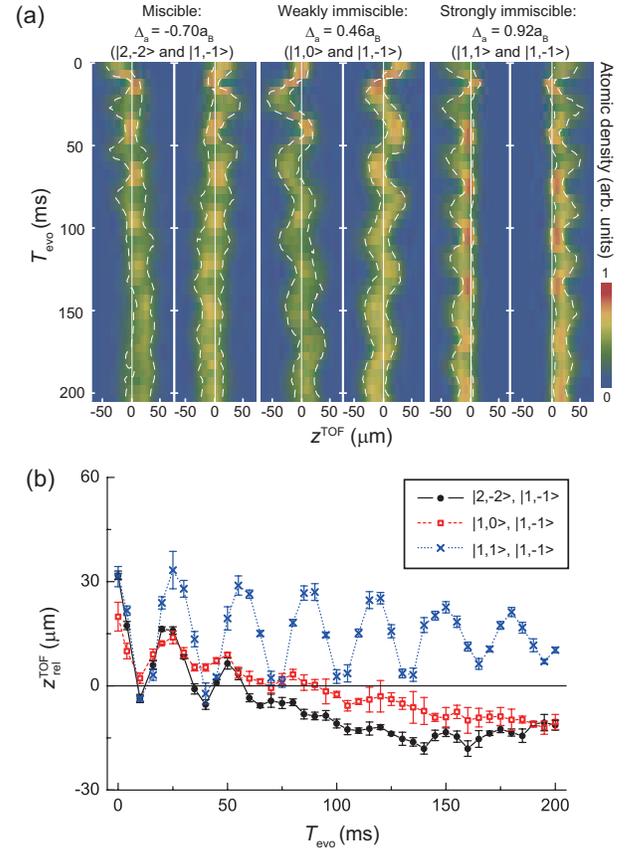}
\caption{
(color online) (a) $T_{\mathrm{evo}}$ versus atomic density distributions for three pairs of binary BECs.
Each distribution corresponds to the average values obtained over three measurements. 
The dotted curves represent contours at 0.7 times the density of each maximum.
(b) $T_{\mathrm{evo}}$ versus the relative center of mass position after the TOF ($z^{\mathrm{TOF}}_{\mathrm{rel}}$).
Each point represents the average values obtained over three measurements, and the error bars indicate the standard deviation of those measurements.  
}
\end{figure}

\begin{figure}[t]
\includegraphics[width=7.5cm]{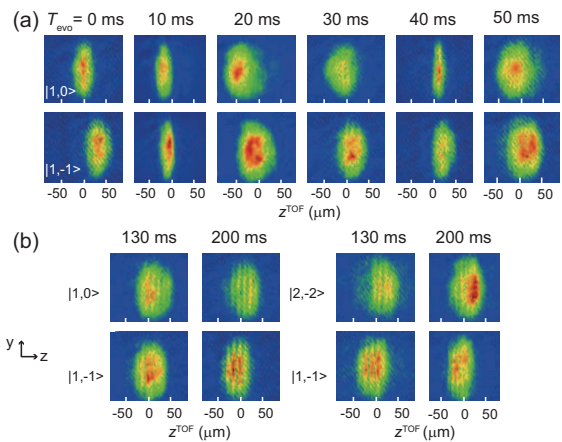}
\caption{
(color online) (a) Absorption images of weakly immiscible BECs at early $T_{\mathrm{evo}}$ values.
(b) Absorption images of weakly immiscible and miscible BECs at $T_{\mathrm{evo}} = 130$ ms and $200$ ms.
These images are obtained from a single shot measurement.
In the density distributions of Fig. 2(a), the domain structures in (b) vanish due to the averaging of the data.
}
\end{figure}

We first investigate the dynamics of colliding binary BECs for three different values of $\Delta_{\rm a}$.
Figure 2(a) shows the $T_{\mathrm{evo}}$ dependence of atomic density distributions for three pairs of binary BECs,
where each distribution corresponds to the average values obtained over three measurements.
In order to extract bouncing and penetrating motions from the data in Fig. 2(a),
relative center of mass positions ($z^{\mathrm{TOF}}_{\mathrm{rel}}$) are calculated, which are shown in Fig. 2(b).
Note that the center of mass positions after the TOF do not correspond exactly to those in the trap, and their values might be affected by the in-trap velocity.
However, in the previous study \cite{Eto15}, we confirmed that the numerically obtained in-trap motion agrees qualitatively with  the experimentally obtained $z^{\mathrm{TOF}}_{\mathrm{rel}}$ values due to the short TOF time.

The common feature in all of the pairs shown in Fig. 2(b) is that $z^{\mathrm{TOF}}_{\mathrm{rel}}$ oscillates for small $T_{\mathrm{evo}}$, namely, the BECs bounce at least once.
The difference appears after a few bounces, and their motions are classified into two categories.
In the case of the strongly immiscible pair, $z^{\mathrm{TOF}}_{\mathrm{rel}}$ oscillates in the range of $z^{\mathrm{TOF}}_{\mathrm{rel}} \geq  0$ $\mu$m, 
where the oscillation frequency of 33 Hz is larger than the axial trap frequency.
This indicates that the strongly immiscible binary BECs continue to bounce without passing through each other.
On the other hand, in the case of the other pairs, $z^{\mathrm{TOF}}_{\mathrm{rel}}$ becomes negative after a few damped oscillations, namely, the two wave packets mutually penetrate after a few bounces.

Although in our experiment the energy given by the magnetic field gradient is one order of magnitude larger than the effective energy barrier \cite{Kurn99} for immiscible BECs to tunnel through each other, 
all of the pairs exhibit the bouncing motion at the first collision.
This seemingly counterintuitive behavior can be understood by considering the energy required to overlap the BECs.
If the spatially separated BECs were overlapped without bouncing,
a large additional energy would be required due to the interspecies mean field interaction proportional to $a_{1,2}$, which is much larger than the effective energy barrier.
The magnitudes of $a_{1,2}$ for the three pairs are almost the same, and thus the bounce occurs even in the miscible system. 

The penetrability of the colliding BECs is sensitive to $\Delta_{\rm a}$, although the energy given by the magnetic field gradient is larger than the effective energy barrier.
This is likely due to the transfer of the energy of the center of mass motion into the spatial structures. 
Figures 3(a) and 3(b) show typical absorption images of the bouncing motion and the penetration process, respectively.
The width in the $z$-direction is significantly changed with the bouncing motion [Fig. 3(a)], and in this process, the domain structure is formed regardless of miscibility [Fig. 3(b)].
These experimental results show that the acquired kinetic energy of the center of mass is converted to the energies of the domain formation and the spatial narrowing.

\begin{figure}[t]
\includegraphics[width=8cm]{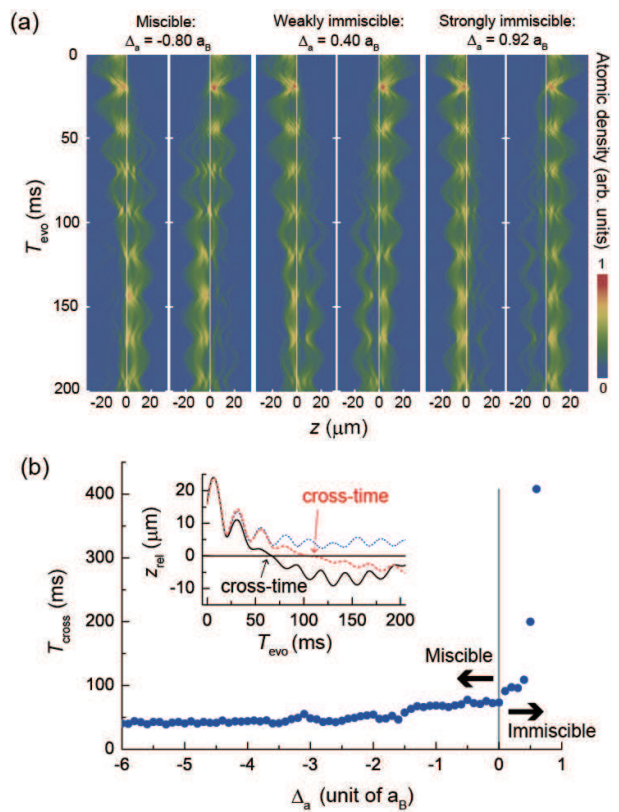}
\caption{
(color online) Numerical simulation of in-trap motion with various degrees of miscibility.
In these numerical simulations, the $a_{1,2}$ values between $\ket{1,1}$ and $\ket{1,-1}$ are varied.
Although we assume a harmonic trap in the numerical simulations, the trap created by the Gaussian beams is shallower in the actual experiments.
This difference leads to the discrepancy in the initial potential energies.
By tuning the value of $T_{\mathrm{sep}}$ in the simulations, we can compensate for this discrepancy. 
(a) $T_{\mathrm{evo}}$ versus atomic density distributions for $\Delta_{\rm a} = -0.8 a_{\rm B}$, $0.4 a_{\rm B}$, and $0.92 a_{\rm B}$
(See the Supplemental Material for movies of the dynamics of the density profile \cite{Supp}).
(b) Cross-time ($T_{\mathrm{cross}}$) as a function of $\Delta_{\rm a}$.
Here $z_{\mathrm{rel}}$ first crosses zero at $T_{\mathrm{cross}}$.
The inset shows $T_{\mathrm{evo}}$ versus $z_{\mathrm{rel}}$,
where solid, broken, and dotted curves represent data for $\Delta_{\rm a} = -0.8 a_{\rm B}$, $0.4 a_{\rm B}$, and $0.92 a_{\rm B}$, respectively.
}
\end{figure}

We numerically simulate the dynamics of colliding binary BECs by changing the value of $a_{1,2}$ between $\ket{1,1}$ and $\ket{1,-1}$.
In order to investigate the relationship between $\Delta_{\rm a}$ and the penetrability,
we calculate the cross-time $T_{\mathrm{cross}}$ at which $z_{\mathrm{rel}}$ first crosses zero [inset of Fig. 4(b)].
Figure 4(b) shows the $\Delta_{\rm a}$ dependence of $T_{\mathrm{cross}}$.
$T_{\mathrm{cross}}$ sharply increases with $\Delta_{\rm a}$ in the region of $\Delta_{\rm a} > 0$, whereas it is nearly constant for $\Delta_{\rm a} < 0$.
When $\Delta_\mathrm{a} \gtrsim a_\mathrm{B}$, $z_{\mathrm{rel}}$ cannot reach zero, i.e., binary BECs continue to bounce without passing through each other.
This threshold varies depending on the energy given by the magnetic field gradient.
This result shows that the penetrability is sensitive to $\Delta_{\rm a}$ and can be tuned by slightly changing the atomic interaction strengths in immiscible BECs \cite{Supp}.

Damped bouncing motion is also observed in simulations of miscible and weakly immiscible BECs, as shown in Fig. 4(a) and the inset of Fig. 4 (b).
Since the binary BECs mutually penetrate while repelling each other, the spatial domain structures are formed, in accordance with which reproduce the experimental results [Fig. 3(b)].
The fringe patterns existing at $T_{\mathrm{evo}} =0$ ms are created during the course of spatial separation of the BECs.

\begin{figure}[tbp]
\includegraphics[width=8cm]{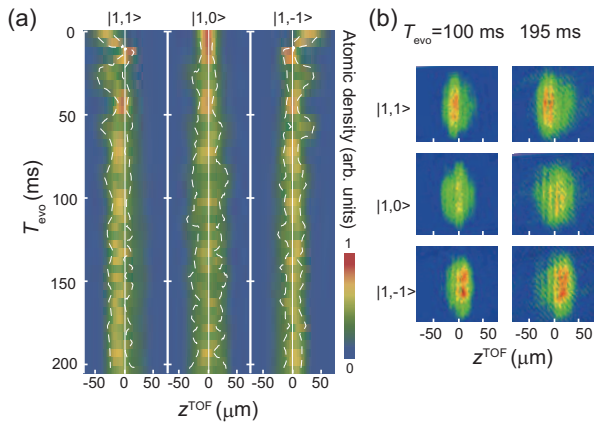}
\caption{
(color online) Time evolution of colliding ternary BECs.
(a) $T_{\mathrm{evo}}$ versus atomic density distributions.
(b) Absorption images of ternary BECs at $T_{\mathrm{evo}} = 100$ ms and $195$ ms.
}
\end{figure}

Finally, we investigate the dynamics of colliding ternary BECs, which are comprised of three spin states in $F = 1$.
Figure 5(a) shows $T_{\mathrm{evo}}$ versus atomic density distributions for the ternary BECs.
The components $\ket{1,1}$ and $\ket{1,-1}$ collide with $\ket{1,0}$.
After a few bounces, the ternary BECs form the domain structures [Fig. 5(b)].
Additionally, in this case, the different components refuse to overlap each other.
Such bouncing motion of multicomponent BECs implies the possibility for realizing a Newton's cradle, which was experimentally observed in a Tonks-Girardeau gas \cite{Kinoshita06} and theoretically predicted for colliding matter-wave solitons \cite{Novoa08}.

In conclusion, we have investigated the dynamic properties of penetration and bouncing motion in colliding binary and ternary BECs in an optical trap.
In contrast to previous experiments on collisions between BECs in a magnetic trap \cite{Maddaloni00, Modugno02},
the use of an optical trap enabled us to generate multicomponent BECs with various degrees of miscibility.   
The observed dynamics are classified into two categories: repeated bouncing motion and mutual penetration after damped bouncing motion.
The various counterintuitive effects such as mutual penetration in immiscible BECs, bouncing between miscible BECs, and domain formation in miscible BECs were observed.
In addition, our numerical simulations show that the properties of penetration and bouncing can be tuned by slightly changing the atomic interaction strengths.
These results suggest the possibility of controlling the dynamics of binary BECs by slightly tuning their interaction strengths.
For example, bouncing and penetration properties in binary BECs respectively enable one component to act as a tunable atom mirror and the other as a dispersive medium. 
Our investigation on the dynamics of multicomponent condensates will be useful in the field of quantum hydrodynamics and atom optics.   
Also our method can be used to study the thermalisation mechanism of an isolated quantum system \cite{Gring12}.

We thank S. Tojo, M. Sadgrove, and Y. Masuyama for their contributions in the early stage of this work.
We also thank M. Tsubota and H. Takeuchi for fruitful discussions.
This work was supported by a Grant-in-Aid for Scientific Research (C) (No. 26400414, No. 15K05233), a Grant-in-Aid for Scientific Research on Innovation Areas Fluctuation \& Structure (No. 25103007) from 
the Ministry of Education, Culture, Sports, Science, and Technology of Japan, and Research Grants of Yoshishige Abe Memorial Fund.

\end{document}